\documentclass[prd,%twocolumn%,preprint
,superscriptaddress,showpacs,nofootinbib,%
tightenlines]{revtex4}
\usepackage{epsfig}
\usepackage{amssymb,amsmath}
\usepackage{ulem}
\newcommand{\beq}{\begin{equation}}
\newcommand{\eeq}{\end{equation}}
\newcommand{\beqa}{\begin{eqnarray}}
\newcommand{\eeqa}{\end{eqnarray}}

\newcommand{\ben}{\begin{displaymath}}
\newcommand{\een}{\end{displaymath}}
\newcommand{\be}{\begin{equation}}
\newcommand{\ee}{\end{equation}}
\newcommand{\bea}{\begin{eqnarray}}
\newcommand{\eea}{\end{eqnarray}}

\usepackage{color}

\begin{document}
\title{%Symmetry-preserving
How (not) to renormalize integral equations with singular potentials \\ in effective field theory}
\author{E.~Epelbaum}
\affiliation
{Institut f\" ur Theoretische Physik II, Fakult\" at f\" ur Physik und Astronomie, \\ Ruhr-Universit\" at Bochum 44780 Bochum, Germany}
\author{A.~M.~Gasparyan}
\affiliation
{Institut f\" ur Theoretische Physik II, Fakult\" at f\" ur Physik und Astronomie, \\ Ruhr-Universit\" at Bochum 44780 Bochum, Germany}
\affiliation
{Institute for Theoretical and Experimental Physics, B. Cheremushkinskaya 25, 117218 Moscow, Russia}
\author{J.~Gegelia}
\affiliation{Institute for Advanced Simulation, Institut f\"ur Kernphysik
   and J\"ulich Center for Hadron Physics, Forschungszentrum J\"ulich, D-52425 J\"ulich,
Germany}
\affiliation{Tbilisi State  University,  0186 Tbilisi,
 Georgia}
\author{Ulf-G.~Mei\ss ner}
\affiliation{Helmholtz Institut f\"ur Strahlen- und Kernphysik and Bethe
   Center for Theoretical Physics, Universit\"at Bonn, D-53115 Bonn, Germany}
 \affiliation{Institute for Advanced Simulation, Institut f\"ur Kernphysik
   and J\"ulich Center for Hadron Physics, Forschungszentrum J\"ulich, D-52425 J\"ulich,
Germany}

\date{6 October, 2018}
\begin{abstract}

We discuss the connection between the perturbative and non-perturbative renormalization and 
related conceptual issues in the few-nucleon sector of the low-energy effective field theory 
of the strong interactions.
General arguments are supported by examples from effective theories with and without pions 
as dynamical degrees of freedom. A quantum mechanical potential with explicitly 
specified short- and long-range parts is considered as an
``underlying fundamental theory'' and the corresponding effective field theory 
potential is constructed.
Further, the problem of the effective field theoretical renormalization of the Skornyakov-Ter-Martyrosian equation 
is revisited.

\end{abstract}

\maketitle

\section{\label{introduction}Introduction}

The underlying idea of the chiral effective field theory (EFT) of Quantum Chromodynamics (QCD) 
is that at low energies expressions of physical quantities given by
QCD can be presented as perturbative expansions in powers of small masses and 
energy/momenta divided by some large scale(s).
The aim of an EFT is to reproduce these perturbative series by applying an organising rule,
the so-called  power counting, based on the most general effective Lagrangian and
requiring that the terms in the EFT Lagrangian up to given order generate contributions 
in physical quantities up to the corresponding order.
In the few-nucleon sector of EFT, pioneered in Refs.~\cite{Weinberg:rz,Weinberg:um},  
the issue of renormalization
turned out to be most problematic, leading to controversial approaches and 
considerations \cite{Harada:2006cw,Ordonez:1992xp,Kaiser:1997mw,Kaiser:1999ff,Bedaque:1999ve,Bedaque:2002yg,
Griesshammer:2005ga,Kaiser:2001pc,
PavonValderrama:2003np,PavonValderrama:2005gu,PavonValderrama:2005uj,
Birse:1998dk,Birse:2003nz,Birse:2005um,Steele:1998zc,Lutz:1999yr,
Higa:2003jk,Higa:2003sz,Cozma:2007hc,Girlanda:2007ds,Epelbaum:2002gb,
Epelbaum:2000mx,Epelbaum:2002vt,Epelbaum:1998ka,Epelbaum:1999dj,Kaplan:1996xu,Kaplan:1998tg,
Kaplan:1998we,Savage:1998vh,Cohen:1999ia,Fleming:1999ee,Beane:2001bc,Nogga:2005hy,Lepage:1997cs,
Park:1998cu,Lepage:1999kt,Yang:2009pn,Birse:2010fj,Valderrama:2009ei,
Valderrama:2011mv,Long:2011xw,Epelbaum:2009sd,Djukanovic:2006mc,Gegelia:2001ev,Gegelia:1998xr,Ren:2016jna,PavonValderrama:2016lqn,Long:2007vp,Beane:2000wh,Frederico:1999ps,Timoteo:2005ia,Timoteo:2010mm}.
 For recent reviews and references see e.g.
Refs.~\cite{Bedaque:2002mn,Epelbaum:2005pn,Epelbaum:2008ga,Machleidt:2011zz,Epelbaum:2012vx,Birse:2009my,Valderrama:2016koj}.
In the few-nucleon sector one deals with non-perturbative expressions of physical quantities 
obtained by solving integral (or differential) equations plagued by ultraviolet (UV) divergences.
The main source of disagreement seems to be a different understanding of the relation 
between perturbative and non-perturbative renormalization.
The aim of the current paper is to bring some clarity to this issue.

Leaving out the whole complexity of the technical details our point of view 
can be summarized by saying that renormalization means expressing physical quantities in terms of
other physical quantities instead of the bare parameters of the Lagrangian \cite{Gasser:1982ap}. 
If all UV divergences disappear from the physical quantities after renormalization such a 
theory is called renormalizable.
Renormalization is perturbative if it is applied to perturbative expressions and 
non-perturbative if applied to non-perturbative ones.

While there might exist quantum field theories (QFTs) in four space-time dimensions which 
are perturbatively non-renormalizable and non-perturbatively finite, a consistent 
EFT Lagrangian includes all of the infinite number of interactions
allowed by the underlying symmetries, and therefore every ultraviolet divergence showing up
in physical quantities is canceled by a corresponding counterterm \cite{Weinberg:mt}.

In the few-nucleon sector of chiral EFT, an infinite number of
Feynman diagrams contribute to the scattering amplitude already at leading order.  
Defining effective potentials as sums of irreducible diagrams one sums
up these infinite sets of diagrams by solving the corresponding integral equations (or
Schr\"odinger equation) \cite{Weinberg:rz,Weinberg:um}. 
 
In EFT, un-subtracted {\it regularized} non-perturbative expressions obtained by 
solving integral equations, when expanded in powers of $\hbar$ (corresponding to the
loop expansion), reproduce the {\it regularized} perturbative series.
Let us emphasize that this is {\it not} to say that non-perturbative expressions 
cannot contain {\it more} than perturbation theory can provide. In general, non-perturbative
expressions of EFT may contain contributions which lead to trivial contributions in 
the perturbative series. A  nice example is given by  the instanton contribution in the action of QCD.
It is given by $\exp(-A/g^2)$, with  $A$  some constant and $g$ the strong coupling constant. 
This non-perturbative contribution has a perturbative expansion  (in powers of $g$) of the
form $0+0+0+ \ldots$, i.e. it is trivial at any order. 
Notice, however, that an EFT is obtained by quantizing the corresponding 
classical theory, i.e. its whole construction is based on the assumption that 
the $\hbar\to 0$ limit coincides with the classical EFT. The authors are not aware of any result 
obtained in EFT which indicates that regularized non-perturbative expressions of physical quantities if
expanded around the $\hbar=0$ limit do not reproduce the perturbative series 
of regularized EFT Feynman diagrams.

As mentioned above, EFTs are renormalizable QFTs. It is very well known how to renormalize 
Feynman diagrams.
In self-consistent EFTs properly renormalized non-perturbative expressions, when 
expanded in $\hbar$, must reproduce the
renormalized perturbative series. There is no reason to expect that this is not the case, 
although, except of some very special cases, in general it is not feasible to 
solve equations in closed forms and to carry out the non-perturbative renormalization explicitly.

Any EFT Lagrangian is written in terms of bare parameters and fields. To carry out the 
QFT renormalization one needs either to split the bare
quantities in renormalized ones and counterterms or, equivalently, apply the 
BPHZ renormalization procedure (see, e.g., Ref.~\cite{Collins:1984xc}).
For  few-body problems, the BPHZ procedure can be applied only in very special cases, 
therefore an essential issue is what are the power counting rules for the EFT 
Lagrangian meant to be applied to?
To operators with bare quantities, or with the renormalized ones?
We advocate the point of view that power counting should be applied to interaction terms 
of the EFT Lagrangian with renormalized coupling constants, {\it not} to bare ones,
unless the Wilsonian approach is used, in which case the cutoff should be kept between the soft and 
the hard scales of the problem   
\cite{Birse:1998dk,Harada:2006cw,Epelbaum:2017tzp}.
The power counting for the terms in the EFT Lagrangian with renormalized coupling 
constants translates into a corresponding power counting for physical quantities.
On the other hand, there is no power counting for counterterms as they are divergent 
in the removed cutoff limit and only make sense in combination with corresponding 
contributions of loop diagrams.

In case of the pionless theory for the nucleon-nucleon (NN) interaction it is straightforward 
to implement the power counting applied to renormalized couplings by using BPHZ renormalization.
In this case the equivalent counterterm formalism can also be worked out exactly.
Within the BPHZ scheme, all loop diagrams are subtracted and the bare couplings are replaced by 
renormalized ones.
In the language of counterterms this means that the power counting is applied to sums 
of unsubtracted loop diagrams and corresponding counterterm diagram(s), {\it but not 
separately to each of them}.
The renormalization procedure becomes much more complicated when pions are 
included as dynamical degrees of freedom. Closed expressions cannot be obtained and, 
therefore,  subtractive renormalization cannot be carried out except for some very 
special cases. Lessons learned from exactly solvable  pionless EFT tell us, at least, 
what procedure should {\it not} be followed in order to carry out a self-consistent EFT 
renormalization and {\it not just to get rid off the ultraviolet divergences.}

Our paper is organized as follows.  In section~\ref{contactintpotential} we consider a 
very well known simple example of the S-wave NN scattering in pionless EFT.
Section~\ref{OPE} addresses the case with the one-pion-exchange (OPE) potential. 
A non-singular potential considered as the ``underlying theory'' will be compared 
to the corresponding leading order (LO) EFT potential in Sec.~\ref{toymodel}.
Sec.~\ref{SSTM} considers the Skornyakov-Ter-Martyrosian equation for the three particle
system  and we summarize our findings in Sec.~\ref{summary}.

\section{\label{contactintpotential}Contact interaction potential}

In this section we consider an exactly solvable  EFT potential of
  contact interactions and recapitulate the problems encountered  (and their
  solutions) in connection with the renormalization of non-perturbative expressions.
This will prove to be useful for the considerations in the
next sections where we deal with the cases in which no exact analytic solution can be obtained.

\begin{figure}[t]
\epsfig{file=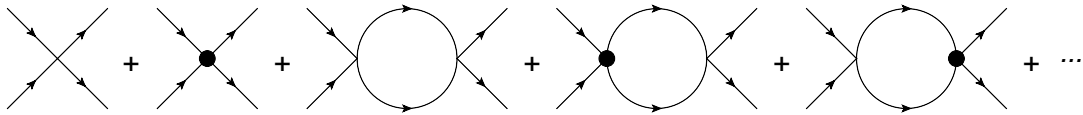,width=0.8\textwidth}
\caption[]{\label{BBs} Diagrams contributing to the NN scattering amplitude. 
The filled blob denotes the NLO  interaction $\sim c_2$. }
\end{figure}

We start by considering the integral equations for the NN
partial wave (PW) scattering amplitudes
\begin{equation}
T^{sj}_{ll'}\left( p,p',q\right)=V^{sj}_{ll'}\left( p,p'\right)+\hbar \sum_{l''} \int_0^\infty \frac{d k \,k^2}{(2\,\pi)^3}\,V^{sj}_{ll''}\left(
p,k\right)\,\frac{m}{q^2-k^2+i\,0^+}\,T^{sj}_{l''l'}\left(
k,p',q\right) \,,
\label{LSequationPW}
\end{equation}
which can be rewritten  symbolically as
\begin{equation}
T=V+\hbar \,V GT~.
\label{one}
\end{equation}
Note that we have included the factor $\hbar$ accompanying the loop integration as the
expansion in $\hbar$ corresponds to the standard QFT loop expansion.
Below we revisit the $^1S_0$ PW in nucleon-nucleon scattering up to next-to-leading order (NLO) 
in pionless EFT considered in Ref.~\cite{Beane:1997pk} and reiterate the
considerations of Refs.~\cite{Gegelia:1998xr,Epelbaum:2009sd,Gegelia:1998gn,Gegelia:1998iu} 
but also adding some new insight.
The starting NLO potential has the form
\begin{equation}
V_{\rm NLO}=c+c_2
\left( p^2+p'^2\right).
\label{1s0}
\end{equation}
This potential is apparently perturbatively non-renormalizable, i.e. its iterations generate 
divergences, which can not be subtracted by redefining the available two parameters $c$ and $c_2$.
On the other hand, in the framework of EFT all divergences generated by iterations of the potential of Eq.~(\ref{1s0}) are systematically removed by counter terms  generated by interaction terms with higher derivatives.  
The on-shell amplitude corresponding to the potential of Eq.~(\ref{1s0}), i.e. the solution to Eq.~(\ref{one}), reads~\cite{Beane:1997pk}:
\begin{equation}
T_{\rm NLO}(q) =\frac{c_2 \left[\hbar\, c_2 \left(I_3
   q^2-I_5\right)-2
   q^2\right]-c}{\hbar \, I\left(q^2\right)
   \left[c_2 \left( \hbar\,c_2
   \left(I_5-I_3 q^2\right)+2
   q^2\right)+c\right]-\left(\hbar I_3
   c_2-1\right){}^2},
\label{1s0ampl1}
\end{equation}
where the integrals $I_n$ are divergent and require regularization.
In cutoff regularization these loop integrals are given by
\begin{eqnarray}
I_n&=& - m\int \frac{d^3k}{(2\,\pi)^3}\ k^{n-3}\,\theta(\Lambda-k)
=-\frac{m\,\Lambda^n}{2\,n\,\pi^2}\,,\nonumber\\
I(p^2)&=& m\int \frac{d^3
k}{(2\,\pi)^3}\,\frac{1}{p^2-k^2+i\,0^+}\,\theta(\Lambda-k)  = -\frac{i\,p\ m_N}{4 \pi}-\frac{m}{2\,\pi^2}\left[ \Lambda
+\frac{p}{2}
\,\ln\frac{1-\frac{p}{\Lambda}}{1+\frac{p}{\Lambda}}\right]\nonumber\\
&& =-\frac{i\,p\,m}{4 \pi}-\frac{m\,\Lambda}{2\,\pi^2} +\frac{m
\,p^2}{2 \,\pi^2 \Lambda}+O\left(\frac{1}{\Lambda^2}\right)
\label{loopintegrals}. \label{integrals}
\end{eqnarray}
\noindent
Our aim is to renormalize the expression of Eq.~(\ref{1s0ampl1})
  in a way consistent with the philosophy of EFT, i.e.~to remove  all
  divergences by absorbing them in the redefinition of the parameters
  of the effective Lagrangian.
For any finite cutoff $\Lambda$ the expansion of $T_{\rm NLO}(q)$ in powers of $\hbar$ is a 
convergent series for sufficiently small $\hbar$ and exactly
coincides with the perturbative series given by perturbative calculations of 
diagrams using the same EFT Lagrangian which has generated the potential of Eq.~(\ref{1s0}).
In particular, the perturbative series
\begin{equation}
T_{\rm NLO}(q)=c+2 c_2 q^2+\hbar \left[c^2 I(q^2)+c_2^2 \left(3 \,I_3
   q^2+ I_5+4 \,I(q^2)\,q^4\right)+2 c_2 c
   \left(I_3+2 \,I(q^2) q^2\right)\right]+\cdots
\label{eqP}
\end{equation}
is in one-to-one correspondence to the diagrams displayed in Fig.~\ref{BBs}.
Renormalization of any of the loop diagrams in Fig.~\ref{BBs} is done according to 
standard textbook  procedures (see e.g. Ref.~\cite{Collins:1984xc}). We can use 
subtractions by applying the BPHZ procedure, which is equivalent to the standard 
counterterm technique. For this particular case, the renormalized series can be easily summed up.  
This subtractive renormalization can be also achieved by replacing the loop integrals 
in Eq.~(\ref{1s0ampl1}) by subtracted ones, e.g., by discarding $I_n$ entirely and 
subtracting $I(q^2)$ at $q^2=-\mu^2$,  
and by simultaneously replacing the bare couplings $c$ and $c_2$ with the
subtraction-scale(s) dependent renormalized ones, denoted $c_R$ and $c_{2R}$, respectively.
The final result reads
\begin{equation}
T_{\rm NLO}(q) =\frac{c_R+2 q^2 c_{2R}}{1- \hbar \, [I (q^2)-I(-\mu^2)] 
\left(c_R+2 q^2 c_{2R}\right)}.
   \label{1s0ampl1R}
\end{equation}
  Notice here that, as usual in any quantum field theory, the
  obtained renormalized result does not depend on the applied
  regularization scheme, e.g.,  
using dimensional regularization and the same renormalization
conditions leads to exactly the same result \cite{Gegelia:1998gn}.
In agreement with our general expectations, the expansion of
Eq.~(\ref{1s0ampl1R}) in $\hbar$ reproduces exactly the renormalized series 
of diagrams. Unlike the result of Ref.~\cite{Beane:1997pk}, the expression in 
Eq.~(\ref{1s0ampl1R})  does {\it not} imply any restriction on the sign of the effective range imposed
by the Wigner bound applicable to a zero-range potential.
The Wigner bound does not apply to the expression of Eq.~(\ref{1s0ampl1R}) because 
its expansion in $\hbar$ generates an infinite series of renormalized diagrams,
i.e.~Eq.~(\ref{1s0ampl1R}) already includes the contributions of an \emph{infinite} number of 
counterterms (with powers of momentum/energy growing up to infinity) and thus 
the corresponding effective potential is not zero-ranged \cite{Gegelia:1998gn}. 
In particular, the bare potential with 
all those counterterms included has the form:
\begin{eqnarray}
V_B &=& \frac{c_R-\hbar c_{2 R}^2 I(-\mu ^2)
   \left(q^2-p^2\right)
   \left(q^2-p'\,^2\right)+c_{2 R}
   \left[p^2+p'\,^2+\hbar c_{2 R} \left(I_3
   \left(p^2+p'\,^2-q^2\right)-I_5\right)\right]}
   {\left(\hbar I_3
   c_{2 R}+1\right){}^2 + \hbar I(-\mu ^2)
   \left[ c_R- c_{2 R} \left(\hbar I_5 c_{2
   R}-q^2 \left(\hbar I_3 c_{2
   R}+2\right)\right)\right]}\nonumber\\
   & = &  c_R+c_{2 R}(p^2 +p'\,^2)  +{\cal O}(\hbar). 
   \label{BareP}
\end{eqnarray}
 This expression is consistent with the standard Weinberg power counting because  
the renormalized potential $c_R+c_{2 R}(p^2 +p'\,^2)$ satisfies this power counting (provided 
that the scale $\mu$ is taken of the order of the hard scale of the problem \cite{Epelbaum:2017byx}) while 
all other terms in Eq.~(\ref{BareP}) are proportional to $\hbar$ - they are counterterms with divergent coefficients which are not subject to the rules of the power counting.
These counterterms contain higher orders of momenta/energy, but that does not cause a problem 
as we are interested in physical quantities and the power counting applies to subtracted loop integrals and  
{\it not} to the diverging loop integrals and diverging counter terms separately.
What is mapped onto the power counting for the physical amplitude is the power counting for 
the potential with renormalized couplings.  Notice that while,  for the case of unnaturally 
large scattering length  and natural effective range, the LO potential $c$ must be iterated to obtain the LO EFT amplitude, 
the $c_2$-term can be treated perturbatively, its first insertion
generates the NLO contribution to the amplitude. 
The first two terms in the expansion of the amplitude $T_{\rm NLO}(q)$ of Eq.~(\ref{1s0ampl1R}) in powers 
of $c_{2 R}$ exactly reproduce the LO and NLO contributions to the renormalized EFT amplitude and 
the higher order terms are all small, as expected.

  In case the NLO potential is treated perturbatively, all
  divergences are removed by the bare potential (expanded in $c_{2R}$),
  which does not contain counterterms with higher orders of momenta/energy:
\begin{equation}
V_{B,{\rm expanded}}=\frac{c_R}{1+\hbar c_R
   I(-\mu ^2)}
   - \frac{2
   \hbar c_{2 R} c_RI_3}{\left[1+\hbar c_R
   I(-\mu
   ^2)\right]{}^2} + \frac{c_{2 R} \left[ p^2+p'\,^2 + \hbar c_R I(-\mu ^2)
   \left(p^2+p'\,^2-2 q^2\right)\right]}{\left[1+\hbar c_R
   I(-\mu
   ^2)\right]{}^2}.
   \label{VctExp}
\end{equation}
As it is seen from the above expressions it does not matter whether we treat the $c_2$ term 
perturbatively or non-perturbatively, the difference in the corresponding expressions for the 
scattering amplitude  is indeed of higher order provided that we renormalize it properly, 
i.e.~that we take into account contributions of all necessary counterterms, or equivalently, 
subtract  UV divergences from {\it all} loop diagrams.

\medskip

To summarize, the expression of Eq.~(\ref{1s0ampl1}) reproduces the perturbative diagrams 
if expanded around $\hbar =0$. That perturbative series
is convergent for arbitrarily large (but finite) cutoff and sufficiently small $\hbar$. 
The corresponding renormalized series is also convergent, and it converges to the expression of  
Eq.~(\ref{1s0ampl1R}).
Thus, the non-perturbative expression, when expanded in $\hbar$, reproduces the perturbative 
series and the corresponding renormalized non-perturbative expression, when expanded in $\hbar$, 
reproduces the renormalized perturbative series.

\medskip

  Next, we discuss a procedure often referred to as
  ``non-perturbative renormalization" which makes the non-perturbative
  expression of Eq.~(\ref{1s0ampl1}) finite,  
but does not qualify as a renormalization of EFT as it does not
match the well established quantum field theoretical perturbative
renormalization in the region where perturbation theory is
applicable. 

In Ref.~\cite{Beane:1997pk} the two parameters  $c$ and $c_2$ have been fixed by demanding that 
the amplitude of Eq.~(\ref{1s0ampl1}) reproduces the scattering length and the effective range.
The obtained amplitude has the form
\begin{equation}
T(q)=\frac{-4 i \pi  a \left[4 a \hbar \Lambda +\pi
   \left(a q^2 \,r_e+2\right)\right]}{m \left[\pi
   \left(a^2 q^3 \,r_e+2 a q-2 i\right)+2 a \hbar
   \Lambda  (a q (2+i q \,r_e)-2 i)\right]}= \frac{-4 \pi /m }{-\frac{1}{a}+\frac{q^2 r_e}{2}-i q} -\frac{\pi ^2 a q^4 r_e^2/m}{\hbar \Lambda \left(-\frac{1}{a}+\frac{q^2 r_e}{2}-i q\right)^2}+{\cal O}\left(\frac{1}{\Lambda^2}\right).
\label{cea}
\end{equation}
This expression happens to be finite in the limit $\Lambda\to \infty$ leading to
\begin{equation}
T(q)=-\frac{4 \pi /m }{-\frac{1}{a}+\frac{q^2 r_e}{2}-i q}~~.
\end{equation}
Notice, however, that the expansion of the expression of Eq.~(\ref{cea}) in powers 
of $\hbar$ leads to:
\begin{equation}
T(q)=\frac{2 \pi  a \left(a q^2 r_e+2\right)}{m}+\hbar\left[\frac{2 a^4  \Lambda  q^4 r_e^2}{m}-\frac{i \pi
   a^2  q \left(a q^2 r_e+2\right){}^2}{m}\right]+\cdots ,
\label{caee}
\end{equation}
which is, again, a convergent series for arbitrarily large but finite $\Lambda$ and sufficiently small 
$\hbar$.
In this expression the term of order $\hbar$ as well as all higher-order terms contain positive 
powers of $\Lambda$. Thus, terms in Eq.~(\ref{caee}) correspond to partially
renormalized diagrams, that means that some positive powers of the cutoff are removed while
others are not. We remark that standard quantum field theoretical renormalization applied to
Feynman diagrams of the perturbative series removes all 
positive powers and logarithms of the cutoff parameter.  Thus the perturbative expansion of the
``nonperturbatively-renormalized" expression of Eq.~(\ref{cea}) does not reproduce the
renormalized perturbative series.

Notice that the closed non-perturbative expression of Eq.~(\ref{cea})
satisfies the {\it necessary} condition imposed on a properly renormalised expression that for 
large values of the cutoff the cutoff-dependent part is suppressed by  inverse powers  
of $\Lambda$. However, it is {\it not}  an EFT renormalized
expression unless one defines EFT renormalization in such a way that
there is a clear mismatch between the renormalization of the perturbative series and a convergent 
sum of this series (for small $\hbar$). 

The expression of Eq.~(\ref{cea}) is affected by the Wigner bound, i.e. the cutoff cannot 
be taken very large unless the effective range is non-positive (otherwise the bare 
parameters $c$ and $c_2$ become complex).
 However, this restriction has nothing to do with EFT  because
 Eq.~(\ref{cea}) is {\it not} 
renormalized from the EFT point of view. 

One might argue that we are addressing an irrelevant issue here as the $c_2$-term is not of LO 
for the NN system and therefore we do not have to include it non-perturbatively.
However, there might exist systems (at least theoretically) for which the $c_2$-term needs to be 
iterated, and most importantly we aim at using the pionless EFT as a ``theoretical laboratory'' 
where we can learn how to deal with problems for which we do not have exact expressions.

According to Weinberg's power counting  in an EFT with pions treated as explicit degrees of freedom,
 the OPE potential is of LO. However, 
from the point of view of the renormalization, if compared to the above problem  of the pionless EFT,
it acts {\it not} like the $c$-term but rather like the $c_2$-term,  
i.e.~it is perturbatively non-renormalizable, see e.g. Ref.~\cite{Savage:1998vh}.

Next let us have a closer look at the issue of renormalization for the OPE potential
making use of the lessons learned in this section.

\section{Including the one-pion-exchange}
\label{OPE}

According to Ref.~\cite{Weinberg:rz}  in EFT with pions and nucleons as dynamical degrees of freedom,
the LO NN potential is given by energy- and momentum-independent contact interaction terms plus 
the OPE potential.
If this power counting rule is applied to the bare quantities, then the full LO potential has the form
\begin{equation}
V_{\rm LO}=V_C+V_\pi ,
\label{VPion}
\end{equation}
where $V_\pi$ is the OPE potential and the contact interaction part $V_C$ contributes only in S-waves.
However, following the approach of Ref.~\cite{Nogga:2005hy}  let us apply ``non-perturbative
renormalization" by including an additional 
single contact interaction term in each attractive triplet PW and fixing them by minimizing the 
cutoff dependence
of the solutions of the integral equation for the corresponding PW amplitudes. 
Below we argue that such a ``non-perturbative renormalization" is an {\it ad hoc} procedure, 
inconsistent with the quantum field theoretical renormalization of EFT.

To calculate the scattering amplitude for our LO potential we need to solve the PW integral 
equations~(\ref{one}) to which we apply the cutoff regularization. For arbitrarily large but 
finite cutoff $\Lambda$ and for sufficiently 
small $\hbar$, the iterations of Eq.~(\ref{one}) generate a perturbative
series which converges to the solution of the equation. That is, for arbitrarily large but finite 
cutoff, the solution of  Eq.~(\ref{one}) can be expanded in a convergent series of $\hbar$ 
around $\hbar=0$. This series exactly reproduces the results for diagrams obtained by
iterating the OPE potential and the contact interaction (if the latter
is non-vanishing for a given PW). Each term in this convergent (for
sufficiently small $\hbar$) series of diagrams can be renormalized by
either using BPHZ subtractions or the counterterm technique. Unlike
the analogous case of the pionless EFT, we do not know how to sum up
the resulting renormalized series. However, a simple counting of
orders of UV divergences of  perturbative loop diagrams makes it clear that  the
single contact interaction term included in the potential in each
spin-triplet attractive channel 
cannot generate all those subtractions of loop diagrams (see e.g. Ref.~\cite{Savage:1998vh}). That is, 
for   a large (but finite) cutoff, independently from the choice of the available single 
contact interaction term, which is tuned (as a function of the cutoff) to minimize the cutoff 
dependence in the non-perturbative solution of the corresponding PW integral equation \cite{Nogga:2005hy},
the expansion of the solution to  Eq.~(\ref{one}) in powers of $\hbar$ for sufficiently small 
$\hbar$ generates a convergent series of terms which are only partially renormalized, they contain 
positive powers of the cutoff.
Therefore, these diagrams do not coincide with perturbatively renormalized diagrams.
That is, analogously to the contact interaction potential of the previous section the
``non-perturbatively renormalized" 
expressions, when expanded, do not reproduce the perturbative series renormalized according to
the rules of quantum field theory. We conclude that also for the 
OPE potential, the ``non-perturbative renormalization" is inconsistent with proper EFT renormalization.

One might be tempted to declare this mismatch between ``non-perturbative renormalization"
and renormalized perturbative series, which is even more obvious for the case
of a repulsive singular interaction, as irrelevant by taking the
cutoff to infinity, in which case the solution to the equation becomes
a non-analytic function of the coupling constant of the OPE potential
(see, e.g., Ref.~\cite{Frank:1971xx}). That is,  the resulting non-perturbative amplitude cannot
be expanded perturbatively in powers of the coupling constant, i.e. it
cannot be expanded in powers of $\hbar$ either. One could claim that
non-perturbative expressions have nothing to do with perturbation
theory because it is a solution in an intrinsically non-perturbative
regime. Such an argument is not valid here because the non-analyticity in the coupling constant of
the OPE potential originates from the singular $1/r^3$ behaviour of
OPE potential for $r\to 0$ \cite{Frank:1971xx}. 
The OPE potential of chiral EFT is  obtained in the low-energy region and
its singular $1/r^3$ behaviour for $r\to 0$ has nothing to do with
either EFT or the real world and the underlying fundamental theory,
QCD,\footnote{Thus, while derived from the chiral EFT and
  therefore relevant for large distances, the $\sim 1/r^3$ part of the
  OPE potential  has nothing to do with the nucleon-nucleon
  interaction at shorter distances, where its singular  behaviour 
  shows up.} which is believed to describe the real world
where the only bound 
state in  the NN system
is the deuteron, while the singular $1/r^3$ behaviour inevitably entails deeply bound states.
Notice further that if we could sum up the  properly renormalized perturbative
series for sufficiently small $\hbar$, the existence of its
unique continuation to the physical value of $\hbar$ would depend on
the choice of the renormalization condition. 

 To bring further insight into the above discussion, we consider below
a quantum mechanical non-singular potential with the long range
part behaving like $1/r^3$ if extended to small values of $r$ while 
dropping the short range parts.

\begin{figure}
\epsfig{file=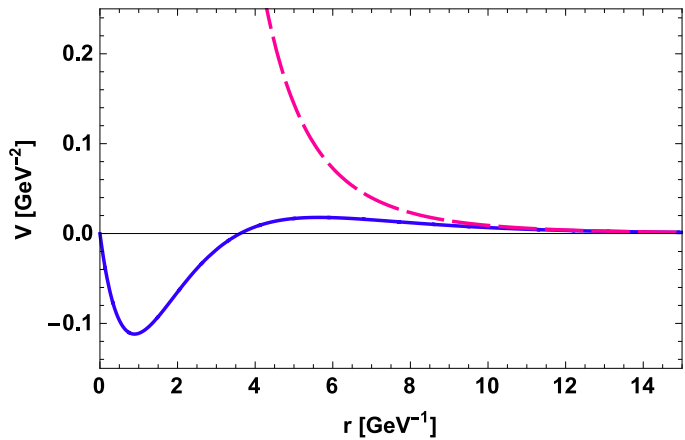,width=0.5\textwidth}
\caption[]{\label{Pots} Exact and approximate potentials as discussed in the text. 
The solid (blue) and the long-dashed (magenta) lines corresponds to the exact and 
the approximate potentials, respectively.}
\end{figure}

\section{Non-singular potential of the ``underlying theory"}
\label{toymodel}

Let us consider the potential
\begin{eqnarray}
V(r) &=& \frac{\alpha  \left(e^{- m_1 r}-e^{-M
   r}\right)}{r^3}+\frac{\alpha  \left(m_1-M\right) e^{- m_1 r}}{r^2}
   +\frac{\alpha  \left(M-m_1\right){}^2 e^{-m_2 r}}{2 r} \nonumber\\
   &-& \frac{1}{6} \alpha  \left(2 m_1-3
   m_2+M\right) \left(M-m_1\right){}^2 e^{- m_1 r},
\label{potentialdef}
\end{eqnarray}
where the light mass $M$ is the small scale and  the heavy masses $m_1$, $m_2$ are the large scales. 
Our choice of parameters is $\alpha=-36\, {\rm GeV^{-2}}$, $M=0.1385$~GeV, $m_1=0.75$~GeV and 
$m_2=1.15$~GeV. The factor $\alpha$  sets the strength of the interaction. This strength is 
taken equal for all terms, so that
the potential $V(r)$ vanishes for $r\to 0$ and it behaves as $-\alpha\, e^{-M r}/r^3$ for large $r$.
In Fig.~\ref{Pots} we show the full potential and its long range part extended to small values of $r$.

\begin{figure}
\epsfig{file=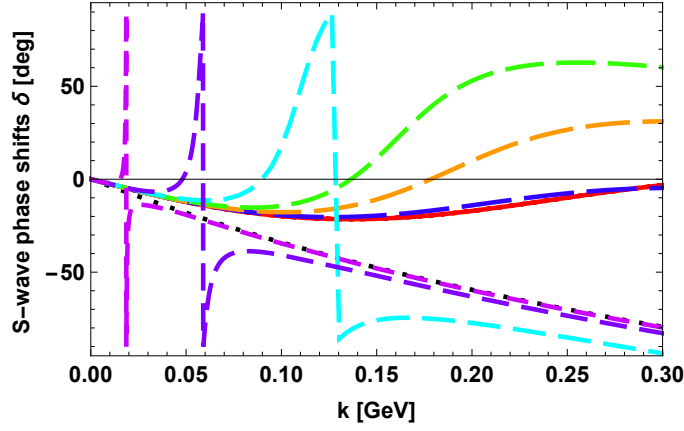,width=0.5\textwidth}
\caption[]{\label{phsR} S-wave phase shifts versus the particle momentum in the center-of-mass frame. 
The solid (red) line corresponds to the underlying potential and the dashed lines with decreasing 
length of dashes are phase shifts for the cutoff $\Lambda = 0.6, 0.8, 1.0, 1.4, 2.0$ and $3.0$~GeV, 
respectively. The constant contact interaction term is fitted to reproduce the scattering length.  
The dotted (black) line represents the phase shifts corresponding to the singular long range 
potential in the infinite cutoff limit.}
\end{figure}

The Fourier transform of the potential $V(r)$ has the form
\begin{eqnarray}
V(q) & = & \int d^3{\bf r}\, e^{i\,{\bf q}\cdot {\bf r}}\,V({\bf r})\nonumber\\
&=& \frac{2}{3} \pi  \alpha  \left(M-m_1\right){}^2
   \left(\frac{3}{m_2^2+q^2}-\frac{2 m_1 \left(2 m_1-3
   m_2+M\right)}{\left(m_1^2+q^2\right){}^2}\right)
   +4 \pi 
   \alpha  \ln \frac{q+i M}{q+i m_1}\nonumber\\
   &+& 
   2 \pi  \alpha  \left(1-\frac{i M}{q}\right) \left[ \ln
   \frac{M+i q}{M-i q}-\ln\frac{m_1+i q}{m_1-i q}\right]\,.
   \label{potentialFTr}
\end{eqnarray}
We consider the LS equation for the scattering amplitude in the center-of-mass frame of two particles 
with unit masses interacting via the potential of Eq.~(\ref{potentialFTr})
\begin{equation}
T_E({\bf p},{\bf p}')=V(|{\bf p}-{\bf p}'|)+\int \frac{d^3 {\bf q}}{(2 \pi)^3}\, V(|{\bf p}-{\bf q}|)\,
\frac{1}{E-q^2+i\epsilon} \, T_E({\bf q},{\bf p}'),
\label{LSEquation}
\end{equation}
with $E=k^2/1 \, {\rm GeV}$  the total energy and ${\bf p}$ and ${\bf p}'$ the relative momenta of 
the incoming and  the outgoing particles, respectively.
We concentrate on the S-wave integral equation which has the form
\begin{equation}
t_E(p,p\,') =  v (p,p\,')+  \int_0^\infty\frac{d q\, q^2}{2\pi^2} \,v(p,q)\,\frac{1}{E-q^2+i\,\epsilon}\,t_E(q,p'),
\label{tEq}
\end{equation}
where $t_E(p,p\,')=(1/2) \int T_E({\bf p},{\bf p}') d \cos \theta_{{\bf pp}'}$ and
 the S-wave potential $v(p,q)= (1/2) \int V(|{\bf p}-{\bf q}|) d \cos \theta_{{\bf p q}}$ is given by
\begin{eqnarray}
v(p,q)&=& -\frac{\pi  \alpha  (p-q)^2 \ln \frac{\left|
   p-q\right| +i M}{\left| p-q\right| +i m_1}}{p q}
   +\frac{\pi  \alpha  \left((p-q)^2-2 i M \left|
   p-q\right| \right) \ln \frac{m_1+i \left|
   p-q\right| }{m_1-i \left| p-q\right| }}{2 p q} \nonumber\\
   &-& \frac{\pi  \alpha  \left((p-q)^2-2 i M \left|
   p-q\right| \right) \ln \frac{M+i \left|
   p-q\right| }{M-i \left| p-q\right| }}{2 p q}
   -\frac{4 \pi  \alpha  m_1 \left(M-m_1\right){}^2
   \left(2 m_1-3 m_2+M\right)}{3
   \left(m_1^2+(p-q)^2\right)
   \left(m_1^2+(p+q)^2\right)}\nonumber\\
   &-& \frac{\pi  \alpha 
   \left(M-m_1\right){}^2 \ln
   \frac{m_2^2+(p-q)^2}{m_2^2+(p+q)^2}}{2 p q} 
   +\frac{\pi  \alpha  (p+q)^2 \ln \frac{i
   M+p+q}{i m_1+p+q}}{p q}
   - \frac{\pi  \alpha  (p+q)
   (p+q -2 i M) \ln \frac{m_1+i (p+q)}{m_1-i
   (p+q)}}{2 p q}\nonumber\\
   &+& \frac{\pi  \alpha  m_1
   \left(m_1-2 M\right) \ln
   \frac{m_1^2+(p-q)^2}{m_1^2+(p+q)^2}}{2 p
   q}+\frac{\pi  \alpha  M^2 \ln
   \frac{M^2+(p-q)^2}{M^2+(p+q)^2}}{2 p
   q}+\frac{\pi  \alpha  (p+q) (p+q-2 i M) \ln
   \frac{M+i (p+q)}{M-i (p+q)}}{2 p q}~.
   \label{VS}
\end{eqnarray}
Notice here that despite its appearance, $v(p,q)$ is a real function of its arguments.
Eq.~(\ref{tEq}) has a unique well-defined solution and the corresponding phase shifts are represented by 
the solid (red) line in Fig.~\ref{phsR}.

Considering the expression of Eq.~(\ref{potentialdef}) as an ``underlying fundamental'' potential, 
we can construct the corresponding EFT.
The LO EFT potential consists of a constant contact interaction corresponding to the delta 
potential in coordinate space and the long range part $ -{\alpha\, e^{-M r}}/{r^3}$, which is 
singular if extended to the small $r$ region.
The coupling constant $\alpha=-36 \ {\rm GeV^{-2}}\approx -1/(0.167 \ {\rm GeV})^2$ is chosen 
such that the full LO potential as well as its long range part are non-perturbative for 
the momenta $k\sim M=0.1385 \ {\rm GeV}$. 
 A simple UV analysis shows that the LO potential is perturbatively non-renormalizable, i.e. 
to remove the divergences from its iterations one needs to introduce counterterms of higher 
order (in momentum and energy) which themselves generate new divergences etc. up to infinity. 
On the other hand, the solution to the regularized integral 
equation has a well defined removed cutoff limit.  However, this
 ``non-perturbatively renormalized'' expression is not only
 conceptually incorrect from the point of view of EFT but also does
 not describe the data as will be seen below.
Analogously to Ref.~\cite{Epelbaum:2014efa} we regularize the LO potential by multiplying 
its singular long-range part with a local regulator  
function $f(r/R)$ which we choose for convenience of the following form
\begin{equation}
f\left(\frac{r}{R}\right)=\left(1-e^{-\frac{r}{R}}\right)^2,
\label{lreg}
\end{equation}
and the constant contact interaction is regularized in momentum space by multiplying with
\begin{equation}
e^{-\frac{p^2+q^2}{\Lambda^2}}~.
\label{Creg}
\end{equation}
We choose $R= 2/\Lambda $ and fix the constant contact interaction term by demanding that the 
scattering length of the underlying potential is reproduced. 
As expected from general consideration of cutoff EFT
\cite{Lepage:1997cs,Gegelia:1998iu,Gegelia:2004pz,Epelbaum:2006pt} for
the cutoffs around the optimal value $\sim 0.6$~GeV (which is of the order of the large 
scale of the problem) the phase shifts are reasonably well described
by the LO EFT potential. For increasing cutoff, the region where the
phase shifts are well described at LO decreases eventually vanishing in the removed cutoff limit.
Note further that the phase 
shifts corresponding to a repulsive long-range singular potential without adding a strong 
attractive contact interaction have a well defined removed cutoff limit which strongly 
deviates from the data as seen in Fig.~\ref{phsR}.  
One might be tempted to try to reproduce the data by treating the higher order contact 
interactions perturbatively. 
However, and on top to the conceptual problems discussed above,
such a perturbative treatment would be questionable due to the large
discrepancy between the data and the LO  phase shifts, see
also Ref.~\cite{Epelbaum:2015sha} for a related discussion. 
Thus, as mentioned above,  taking the cutoff beyond the large scale of
 the problem for the
  considered perturbatively non-renormalizable LO potential (without including contributions of an infinite number of counter terms) is not
  only  conceptually incorrect from the point of view of EFT \cite{Epelbaum:2009sd}, 
 but does not allow one to describe the data either.

\section{Renormalizing the Skornyakov-ter-Martirosyan equation in EFT}
\label{SSTM}

The arguments of the previous sections also apply to the doublet channel neutron-deuteron 
scattering in pionless EFT as well as the three-body problem of self-interacting scalar 
particles as the corresponding
LO integral equations have characteristic features of  a singular $1/r^2$ potential.

Consider the integral equation for the half off-shell amplitude of the bound state of two 
scalar particles scattering off another scalar particle
\begin{eqnarray}
t(p,k)  &=& M(p,k)+V_3(\Lambda,E,p,k)
+  \hbar \,\frac{2}{\pi} \int_0^\infty \frac{d q\, q^2 t(q,k) }{-\frac{1}{a_2}
+\sqrt{3 q^2/4-mE}}\,F(k,q,\Lambda)
\left[M(p,q)+V_3(\Lambda,E,p,q)\right] ,\nonumber\\
M(k,p) & = &  \frac{1}{p k}\ln \frac{k^2+p^2+k p -m E}{k^2+p^2-k p -m E}~~.
\label{STM}
\end{eqnarray}
Here, $E=3 k^2/4 m-B_2$ is the energy of the three-body system,  $B=1/(m a_2^2)$ is the binding energy 
of two scalars with $a_2$ the two-body scattering length, and $F(k,q,\Lambda)$ is an ultraviolet 
regulator, with $\Lambda$  the pertinent cutoff parameter.
We included  a more general three-body interaction $V_3$ than considered in Ref.~\cite{Bedaque:1998km}, 
where this equation has been obtained using a non-relativistic EFT of self-interacting scalar particles.
Eq.~(\ref{STM}) for $V_3=0$ is equivalent
to the Skornyakov-Ter-Martirosyan  (S-TM) equation~\cite{skornyakov}.
For $V_3=0$ and $F(k,p,\Lambda)\equiv 1$ it does not have a unique solution~\cite{danilov}.  
By considering the 
cutoff-regularized S-TM equation one obtains a unique solution, however,  the limit of the 
cutoff going to infinity 
does not exist \cite{Bedaque:1998kg}.
This problem has been solved by considering  a momentum- and energy-independent three-body force 
$V_3(k,p,\Lambda)=H(\Lambda)/\Lambda^2$ and by choosing $H(\Lambda)$ as a function
of the cutoff so that the strong cutoff dependence of the scattering amplitude is canceled. 
The obtained three-body force $H(\Lambda)/\Lambda^2$ exhibits a limit-cycle behaviour as a 
function of the cutoff~\cite{Bedaque:1998km,Bedaque:1998kg}.

As detailed in Ref.~\cite{Epelbaum:2016ffd}  a proper EFT renormalization of the solution 
to Eq.~(\ref{STM}) requires  the inclusion of contributions of an infinite number of local three-body
counterterms. This disagrees with the widely accepted conclusions of
Refs.~\cite{Bedaque:1998km,Bedaque:1998kg}  that a single constant three-body
interaction term is sufficient.  Note that a simple power counting
demonstrates that iterations of  Eq.~(\ref{STM}) with a constant $V_3$
generate overall divergences which cannot be removed by renormalizing
the available constant three-body term but rather require the
inclusion of contributions of higher-order (up to infinity) counterterms. 
Analogously to the previous sections, the non-perturbative
renormalization utilising a single counterterm is not consistent with
the perturbative renormalization. It can be easily seen by considering
the subset of iterated diagrams involving only  a constant three-body term that the 
perturbative expansion of the solution to the equation for small $\hbar$ does
not reproduce the perturbatively renormalized series. 
Indeed, for $V_3(k,p,\Lambda)=v_3(\Lambda)$,  all these diagrams are generated by expanding the
following expression in powers of $\hbar$:
\begin{eqnarray}
t_3(E,\Lambda)  &=& \frac{v_3(\Lambda)}{1-  \hbar \,\frac{2}{\pi} v_3(\Lambda) \int_0^\infty \frac{d q\, q^2}{-\frac{1}{a_2}
+\sqrt{3 q^2/4-mE}}\,F(k,q,\Lambda)}  \nonumber \\
& \equiv &  \frac{1}{1/v_3(\Lambda)-  \hbar \,\frac{2}{\pi} \left( \frac{\Lambda^2}{\sqrt{3}}
+\frac{4 \Lambda}{3 a_2}-\frac{4 \left(m a_2^2 E+2\right) \ln \left(a_2 \Lambda \right)}{3 \sqrt{3} a_2^2} +{\rm finite}\right)}
=\frac{1}{1/v_3^R +  \hbar \,\frac{2}{\pi} \left( \frac{4 m a_2^2 E  \ln \left(a_2 \Lambda \right)}{3 \sqrt{3} a_2^2} +{\rm finite}\right)}~,
\label{t3}
\end{eqnarray}
where we have used a sharp cutoff. Divergences with energy-independent coefficients are absorbed
in the redefinition of the constant three-body force, however, the logarithmic divergence with an
energy-dependent coefficient remains. Expansion of the resulting partially renormalized expression
in $\hbar$ generates partially renormalized loop diagrams, i.e. does not reproduce the
renormalized expressions of diagrams  involving only  a constant three-body term.

On the other hand, using the RG equation we 
are able to calculate the exact three-body force which removes the full cutoff dependence. 
It is no surprise that no limit cycle is observed in the exact three-body force for large 
values of the cutoff. Notice that we obtain the limit cycle corresponding to the Efimov 
physics by taking larger and larger values for the two-body  scattering length $a_2$, in 
agreement with Refs.~\cite{Moroz:2008fy,Harada:2012na}.
Calculation of  the exact three-body force which removes the full cutoff dependence
can be performed as demonstrated below.

\begin{figure}
\epsfig{file=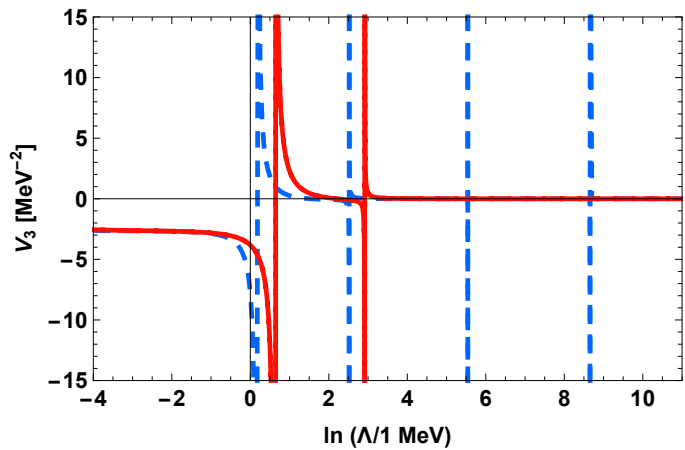,width=0.5\textwidth}
\caption[]{\label{Pot3B} Three-body forces as function of the cutoff.
  The solid and dashed lines correspond to the exact potential and the
  constant contact interaction, respectively. }
\end{figure}

To perform a RG analysis of the S-TM equation~(\ref{STM}), it is convenient to employ the 
regularized Green's function\footnote{Notice here that the 
results of Refs.~\cite{Bedaque:1998km,Bedaque:1998kg} do not depend on the particular form 
of the regulator.} 
\begin{equation}
G_\Lambda(q,k)= \hbar \,\frac{2}{\pi} \,  \frac{q^2 }{-\frac{1}{a_2}+\sqrt{3 q^2/4-mE}}\, 
\left(\frac{k^2+\Lambda^2}{q^2+\Lambda^2}\right)^3\,.
\label{Rgf}
\end{equation}
By demanding  cutoff-independence of the scattering amplitude at  threshold, we obtain the cutoff 
dependence of the constant contact interaction term
$V_3=H(\Lambda)/\Lambda^2$.  We fix the single parameter $H(\Lambda)$ such that $\cot\delta(E) 
= -51.6267$ for $E=-0.00106486$~MeV.
There is no particular reason for choosing these numbers except that they correspond to an 
exactly vanishing three-body force  for $\Lambda=\Lambda_0\equiv 100$ MeV within the given 
regularization scheme.
In all calculations, we take $m=939$~MeV and $a_2=1 \ {\rm MeV}^{-1}$, again for no particular reason.  
As seen from our results plotted in Fig.~\ref{Pot3B}, in full agreement with 
Refs.~\cite{Bedaque:1998km,Bedaque:1998kg} the constant three-body
force shows a limit cycle behaviour for the cutoff getting large. Notice here that we 
plot $H(\Lambda)/\Lambda^2$, not $H(\Lambda)$ as is done in those papers.

Next, by solving the exact RG equation we obtain $V^\Lambda(E,p,k)=M(p,k)+V_3(\Lambda,E,p,k)$, 
which leads to an exactly cutoff independent solution of Eq.~(\ref{STM}).  We fix the 
renormalization conditions of the three-body force by demanding that $V^{\Lambda_0}(E,p,k)=M(p,k)$, 
i.e. the three-body force vanishes for $\Lambda=\Lambda_0$. The exact RG equation for 
$V^\Lambda(E,p,k)$ satisfying the above renormalization condition has the form
\begin{equation}
V^\Lambda=M+M \left(G_{\Lambda_0}-G_\Lambda\right) V^\Lambda.
\label{RGSTM}
\end{equation}
Exact cutoff independence of the  scattering amplitude corresponding to the solution of 
Eq.~(\ref{RGSTM}) can easily be seen by re-writing Eq.~(\ref{STM}) formally as
\begin{equation}
\frac{1}{V^\Lambda}=\frac{1}{t}+G_\Lambda ,
\label{eqinv}
\end{equation}
and subtracting its analogue for $\Lambda=\Lambda_0.$

The three-body force,
i.e. $V_3(\Lambda,E,p,k)=V^\Lambda(E,p,k)-M(p,k)$, is  obtained by
solving numerically Eq.~(\ref{RGSTM}) for  on-shell kinematics with $E=-0.00106486$~MeV and 
is plotted in Fig.~\ref{Pot3B}. 
As seen from the figure no limit cycle is obtained for large values of
the cutoff.

\begin{figure}
\epsfig{file=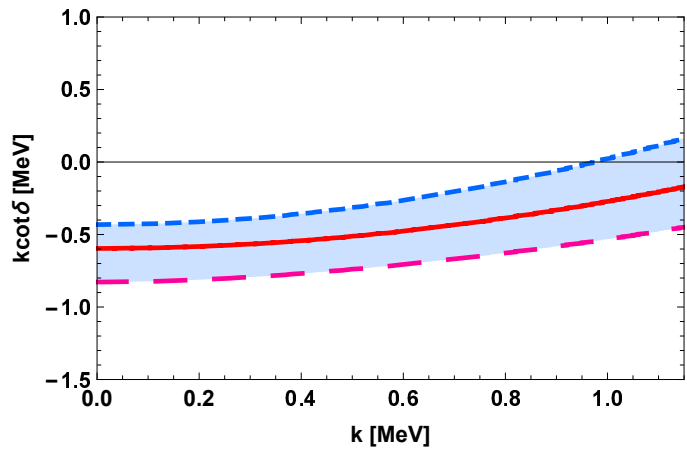,width=0.5\textwidth}
\caption[]{\label{pcot} $k \cot\delta$ as the function of $k$
  calculated without including the three-body force. The short-dashed,
  solid and long dashed lines correspond to $\Lambda=75,\,100,$ and
  125~MeV, respectively.  
  }
\end{figure}

A self-consistent and practically applicable solution to the problem
of non-perturbative renormalization is provided by the cutoff EFT. For
the considered unphysical case the optimal choice is per construction
$\Lambda\sim $~100~MeV. 
We  plot $k\cot\delta$ as the function of $k$ in Fig.~\ref{pcot} for
$\Lambda$ varying between 75 and 125~MeV without including the
three-body force.  As seen from this figure, a 25~\% deviation
from the optimal value of the cutoff  
leads to about a 30~\% change in the physical quantity which is
acceptable. One could reduce the cutoff dependence by promoting the
constant three-body force to the LO as done in
Refs.~\cite{Bedaque:1998km,Bedaque:1998kg},  however, one still must
keep the cutoff in the vicinity of the optimal value $\sim$~hard scale
of the problem to arrive at a self-consistent EFT framework. 

\section{Summary}
\label{summary}

In this work we discussed the connection between the perturbative and non-perturbative 
renormalization in the few-nucleon sector of the low-energy effective field theory of the 
strong interactions.  The corresponding integral equations for the
  scattering amplitude are obtained by means of summing up an infinite
  number of ladder diagrams. Solutions to regularized equations
  reproduce perturbative series of Feynman diagrams when 
expanded in powers of $\hbar$, which corresponds to the loop
expansion. These series are convergent for a finite cutoff and
sufficiently small $\hbar$.
We argued that  properly renormalized non-perturbative physical quantities, 
expanded in powers of $\hbar$, must reproduce
the corresponding renormalized series of standard perturbation theory.
We considered once again the renormalization of the NLO $^1S_0$ nucleon-nucleon potential in 
pionless EFT.  For this exactly solvable model we carried out
  renormalization explicitly and demonstrated the above mentioned
  connection between non-perturbative  
and perturbative expressions for physical quantities. We have also
identified, once again, problems of the so called ``non-perturbative renormalization". 
Next we applied the knowledge gained from this ``theoretical laboratory'' to the 
LO contact interaction and the OPE potential in the pionful EFT.
We argued that the ``non-perturbative renormalization" of the solution to Lippmann-Schwinger 
equation performed by eliminating/minimizing the cutoff dependence of partial wave amplitudes 
for the values of the cutoff much larger than
the hard scale of the problem by adjusting a finite number of counterterms is not consistent 
with  QFT renormalization. To demonstrate our point of view with a further example we considered
 a quantum mechanical non-singular potential with explicitly specified short- and long-range parts as an
``underlying theory'' and addressed the problem of renormalization in the corresponding 
low-energy effective theory. 
We also considered the Skornyakov-Ter-Martirosyan equation describing
the scattering of three scalar particles. We argued that fixing the
non-uniqueness of the solution of this equation by introducing a
single constant three-body force is incompatible with the EFT
formalism. This is because the iterations of the three-body force
generate divergences whose removal requires the introduction of an infinite number of 
other momentum- (and energy-)dependent three-body terms.   
We compared the Wilsonian renormalization group behaviour of the cutoff regulated 
exact three-body force to the cutoff-dependence of the constant three-body force. 
The cutoff dependence of the latter was obtained by demanding cutoff-independence of 
the scattering amplitude at threshold.
  We observe that the constant three-body force as a function of 
the cutoff shows a qualitatively different behaviour compared to the cutoff-dependent 
exact potential for the cutoff much larger than the hard scale of the problem.
In particular, the constant contact interaction exhibits a limit cycle behaviour 
which is absent in the exact potential. The examples considered here support once again 
the arguments of Refs.~\cite{Lepage:1997cs,Gegelia:1998iu,Gegelia:2004pz,Epelbaum:2006pt}
that in the absence of  a possibility of subtracting all UV
divergences  in the solutions of the integral equations of the
non-relativistic EFT, the solution of the problem of renormalization
is provided by cutoff regularized EFT with the cutoff kept of the
 order of the hard scale of the problem.

\section*{Acknowledgments}

This work was supported in part by BMBF (contract No.~05P2015 - NUSTAR
R\&D), by DFG and NSFC through funds provided to the
Sino-German CRC 110 ``Symmetries and the Emergence of Structure in QCD" (NSFC
Grant No.~11621131001, DFG Grant No.~TRR110), by the Georgian Shota Rustaveli National
Science Foundation (grant FR/417/6-100/14), by VolkswagenStiftung (grant no. 93562)
 and by the CAS President's International
Fellowship Initiative (PIFI) (Grant No.~2018DM0034).

\end{document}